\documentclass[letterpaper,10pt]{article} 

\usepackage{osameet3} 

\usepackage{amsmath,amssymb}
\usepackage[colorlinks=true,bookmarks=false,citecolor=blue,urlcolor=blue]{hyperref}
\usepackage{pgfplots}
\pgfplotsset{compat=newest}
\pgfplotsset{plot coordinates/math parser=false}
\newlength{\wlength} \newlength{\hlength}
\pgfplotsset{every tick label/.append style={font=\small}}
\usepackage{graphicx}
\graphicspath{ {images/} }

\begin{document}

\title{Efficient Training of Volterra Series-Based Pre-distortion Filter Using Neural Networks}
\vspace{-3mm}
\author{Vinod Bajaj\textsuperscript{1,2}, Mathieu Chagnon\textsuperscript{2}, Sander Wahls\textsuperscript{1} and Vahid Aref\textsuperscript{2}}
\vspace{-1.7mm}
\address{\textsuperscript{1} Delft Center for Systems and Control, Delft University of Technology, 2628 CN Delft, Netherlands\\
\textsuperscript{2} Nokia Bell Labs, Lorenzstr. 10, 70435 Stuttgart, Germany
}
\email{v.bajaj-1@tudelft.nl, firstname.lastname@nokia.com and s.wahls@tudelft.nl}

\copyrightyear{2022}

\vspace*{-2mm}
\begin{abstract}
We present a simple, efficient 
``direct learning'' approach to train Volterra series-based digital pre-distortion filters using neural networks. We show its superior performance over conventional training methods 
using a 64-QAM 64~GBaud simulated transmitter with varying transmitter nonlinearity and noisy conditions.
\end{abstract}
\section{Introduction}
Digital pre-distortion (DPD) is a key element of the transmitter digital signal processing specially for short-reach applications where the transceiver impairments are more limiting factors. DPD aims to maximize the performance
by compensating the non-ideal response of the transmitter components. Various DPD methods have been so far investigated for 
optical coherent transmitters, e.g. \cite{NAPOLI201852, Yoffe, BajajECOC2020, Paryanti, Faig}. Among them, Volterra series-based DPD methods are quite popular because they are simple to implement and their nonlinear terms are easy to interpret.

A DPD method is usually trained iteratively using either a ``direct learning'' architecture (DLA)~\cite{Eun} or an ``indirect learning'' architecture (ILA)~\cite{Lim}. In ILA, the DPD is trained first as post-equalizer but then it is applied for DPD. 
Although ILA is straightforward, it has limited performance when the nonlinear response is strong or the SNR is low~\cite{Morgan}. DLA does not have these limitations as it learns the pre-inverse but it is more complex to train.
It requires a differentiable surrogate model that approximates the nonlinear response of the system well. An immediate option is to use another Volterra series as surrogate model. However, this option did not get much attention in the past as ``passing'' gradients through a general and flexible model is complex to implement. By applying some approximation in the gradient computation, a simplified version called ``nonlinear filtered-x LMS'' was proposed~\cite{Lim} but this method is still not so easy to use and is penalized by approximations.
A surrogate model should be flexible for extension, have generalization capability over the training sequence, and be easy to use for ``back propagation'' of gradients. Neural networks (NN) have all these properties. They are flexible and suitable for back-propagation due to their modular and multi-layered structure.
They are also well-known for their generalization capability. 
Note that NNs can be also used for DPD, e.g. ~\cite{BajajECOC2020, Paryanti}, but these DPDs are more difficult to interpret and can be computationally complex. Thus, a simpler DPD based on Volterra series can be preferable.
In this paper, we propose to take the best from the both i.e. suitability of NNs to serve as surrogate model and simplicity of Volterra series-based DPD. We show that this new architecture is quite powerful.
We compare its performance with the same Volterra-DPD trained with ILA and with an optimized linear DPD (as a reference of nonlinearity) in simulation.
The considerable SNR gains of this new architecture is quantified for different nonlinearity responses of transmitter.

\section{Simulation setup}
\begin{figure}[!t]
\centering
\includegraphics[scale=0.75]{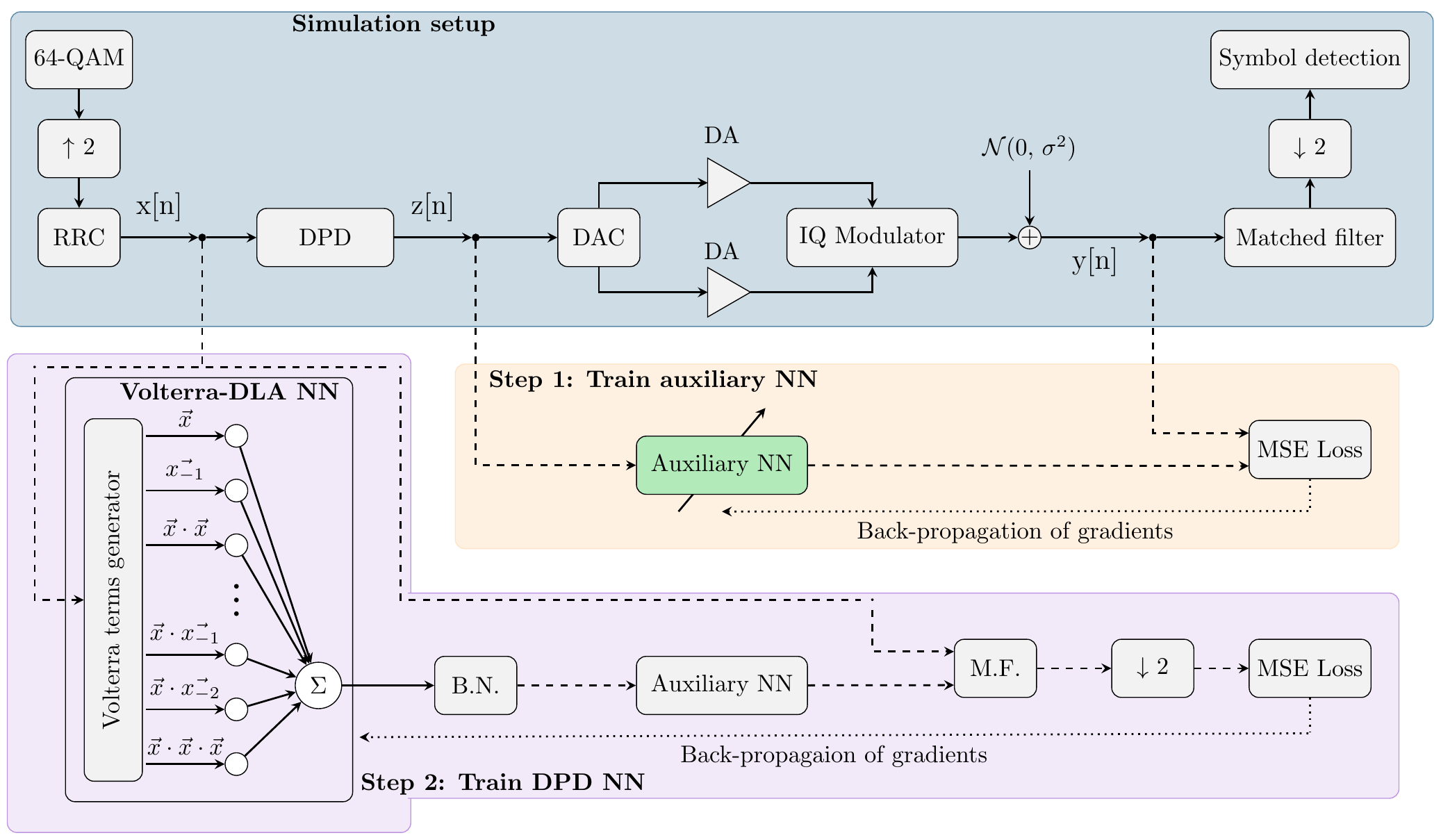}
\vspace{-0.3cm}
  \caption{\small A schematic of 64 GBaud 64-QAM simulation setup (top). The training of the DPD using direct learning architecture (DLA): step 1: training of auxiliary NN (middle) and step 2: training of DPD NN with help of auxiliary NN (bottom). Architecture of the Volterra-DLA NN (bottom-left)}
  \label{fig: setup}
  \vspace{-0.8cm}
\end{figure}

A coherent optical transmitter is modelled as a cascade of digital to analog converters (DACs), driver amplifiers (DAs), a Mach-Zehnder modulator (MZM) and an additive white Gaussian noise (AWGN) source as shown in Fig.~\ref{fig: setup}.
The DACs and the  DAs are simulated as cascades of linear and nonlinear responses, based on the measured linear response of the experimental setup reported in~\cite{BajajECOC2020}. Additionally,
the DAC is impaired by ideal 8-bits quantization and the nonlinear response of the DAs is modelled using the Rapp model~\cite{Paryanti}
\begin{equation}
    v_{\rm out} = v_{\rm in}/\sqrt[4]{ 1 + \left(v_{\rm in}/v_{\rm sat} \right)^{4}}
\end{equation}
where, $v_{\rm in}$, $v_{\rm out}$, $v_{\rm sat}$ are the input, the output and the saturation voltage of the DA.
The nonlinearity of the DA can be controlled by the back-off parameter~\cite{Paryanti}, $\text{back-off [dB]} = 20 \log_{10}(v_{\rm rms}/v_{\rm sat})$
in which $v_{\rm rms}$ is the root mean square value of the input voltage. This back-off parameter definition can also be applied for the DACs and MZM where $v_{\rm sat}$ will be the full scale voltage for DACs or the operating range for the MZM. The back-off values of the DACs and MZMs were fixed to 10 dB and 20 dB, respectively, in order to avoid strong clipping at the DACs and nonlinearity from the MZM. The MZM is modelled as a sinusoidal response, as given in~\cite{Paryanti} with a gain and a phase imbalance of 1\% and $1^\circ$ respectively. 
The transmit signal is a 2sps upsampled 64-QAM signal, shaped using a 64-taps root raised cosine (RRC) filter of roll-off 0.25. The pulse-shaped signal is pre-distored by applying any of the considered DPDs and fed to the DACs. The noisy output signal from the transmitter is then matched filtered and downsampled to 1 sps to extract symbols at the receiver. Later, the normalized mean square error (NMSE) and generalized mutual information (GMI) between the received and the transmit symbols is measured.

We considered three different DPDs in our study: a linear filter with 121 taps memory, Volterra series-based DPDs trained using ILA and DLA, referred next as Volterra-ILA and Volterra-DLA, respectively. A Volterra series up to fifth order was considered. The first order kernel has 121 taps memory. The third order kernels have memory of 10 and depth of 4. The second, fourth and fifth order kernels have memory of 3 and depth 0. 
The linear DPD and Volterra-ILA were trained using ILA using a least mean square approach. The training of the Volterra-DLA is explained in the next section along with its implementation in the NN framework.

\section{DLA Volterra-based DPD and Simulation Results}
The Volterra series-based DPD was designed as a feature mapper (Volterra terms generator) followed by a single layer NN, with single output neuron without any activation, as shown in Fig.~\ref{fig: setup}. The feature mapper takes the real or the imaginary part of the pulse shaped signal as input and maps it into all possible Volterra series terms. The NN takes these Volterra terms as input and produces a predistorted output. The feature length (i.e. the number of neurons in the first layer of the NN) is defined by the Volterra kernels. The Volterra kernels has the same size as described in the previous section. Two disjoint feature mappers and NNs implements the Volterra-DLA DPD for in-phase and quadrature-phase tributary. A batch normalization is performed at the output of the NN. The DPD NN was trained by using another NN which serves as surrogate/auxiliary channel, referred hereafter as auxiliary NN.
We used our previously proposed architecture for the auxiliary NN model~\cite{BajajECOC2020}, but, with the lengths of the first and the last convolution layers changed to 70 and 40 respectively. 

The training of the Volterra-DLA DPD and the auxiliary NN was done in two steps as shown in Fig.~\ref{fig: setup} and also explained in~\cite{BajajECOC2020}. In the first step, the auxiliary NN was trained by using the pre-distorted signal $z$ and noisy received signal $y$ to minimize the mean square error loss between the upsampled waveform. In the second step, the cascaded NN (DPD followed by the auxiliary) is trained, where, only the DPD NN is trainable and the auxiliary is kept fixed. The pulse-shaped signal $x$ and the transmit symbols were used for training in the second step. We have also implemented a matched filtering layer and downsampling layer in the loss computation such that the task is to minimize the mean square error (MSE) loss between the symbols instead of upsampled waveform.

\begin{figure}[!t]
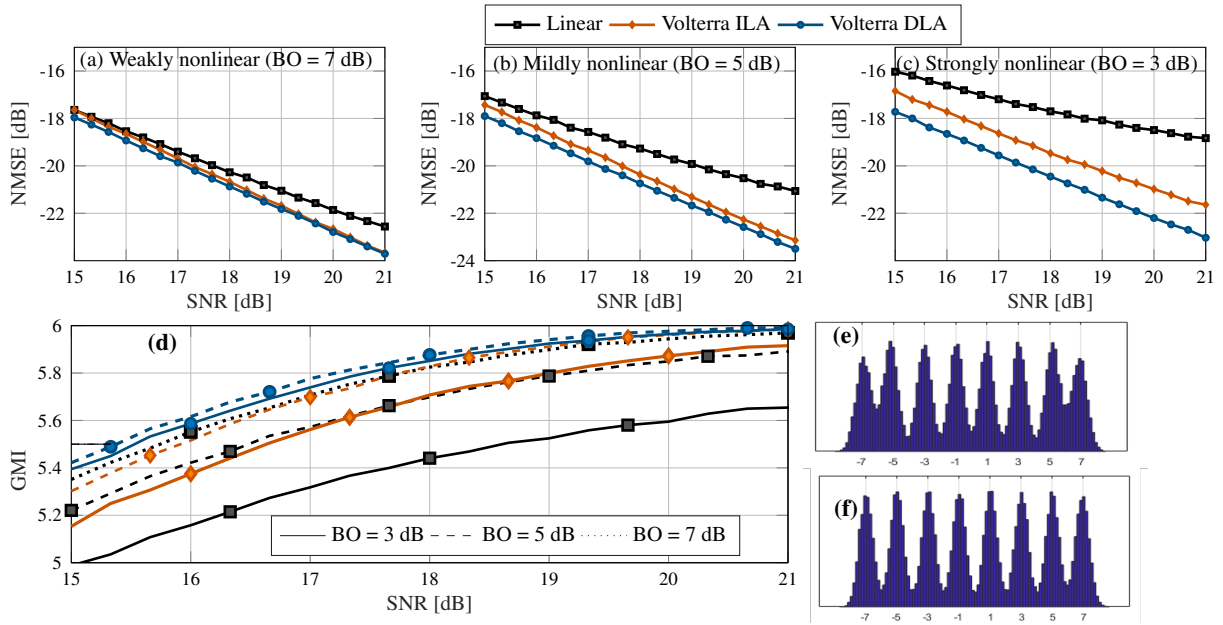

  \begin{minipage}[b]{0.65\textwidth}
\input{GMI_64qam_all}  
  \end{minipage}
  \begin{minipage}[b]{0.35\textwidth}
    \input{histograms}
  \end{minipage}
  \caption{\small (a)-(c) NMSE of the received signal vs channel SNR for different nonlinearity of the DAs. (d) GMI vs. SNR for different back-off values. The colors and markers are the same in all plots. Histograms of the real part    of the received symbols (e) Volterra-ILA (f) Volterra-DLA.}
\vspace{-0.6cm}  
 \label{fig:snr_vs_gmi_16qam}
\end{figure}

The back-off parameter of the driver amplifiers were varied to 7 dB, 5 dB and 3 dB respectively in order to create scenarios with weak, mild and strong nonlinearity. At each back-off value, the SNR of the AWGN source is varied and all the considered DPDs were trained.
Fig.~\ref{fig:snr_vs_gmi_16qam} shows the NMSE values achieved by different DPDs. We see that when the transmitter is weakly nonlinear (back-off=7 dB), the penalty due to nonlinearity is not significant. The linear DPD achieves an NMSE of around -20.3 dB at 18 dB SNR. The nonlinear DPDs, Volterra-ILA and Volterra-DLA achieves 0.4 dB and 0.6 dB smaller NMSE than the linear DPD, respectively. 
In the mildly nonlinear regime (back-off = 5dB), the NMSE with linear DPD degrades by 1 dB to -19.3 dB at 18 dB SNR. The NMSE with Volterra-ILA and Volterra-DLA degrades by 0.4 dB and 0.1 dB from the weakly nonlinear case at 18 dB SNR, achieving -20.4 dB and -20.8 dB NMSE, respectively.
In the highly nonlinear regime (back-off = 3dB), the difference in the performance of the Volterra-ILA and the Volterra-DLA increases. The Volterra-ILA gives -19.5 dB NMSE, while Volterra-DLA gives -20.5 dB NMSE (1 dB less than Volterra-ILA). The linear DPD results in an NMSE of -17.7 dB. When comparing the strongly nonlinear case with the mildly nonlinear case at 18 dB SNR, we see that the NMSE degradation is 1.6 dB, 0.9 dB and 0.3 dB respectively for the linear, Volterra-ILA and Volterra-DLA DPD.

\begin{figure}[!t]
  \begin{minipage}[b]{0.33\textwidth}
\definecolor{mycolor1}{rgb}{0.31373,0.31373,0.31373}%
\definecolor{mycolor2}{rgb}{0.00000,0.44706,0.74118}%
\definecolor{mycolor3}{rgb}{0.72941,0.83137,0.95686}%
\definecolor{myblack}{rgb}{0.31373,0.31373,0.31373}%
\definecolor{mybluef}{rgb}{0.00000,0.44706,0.74118}%
\definecolor{myblue}{rgb}{0.00000,0.30706,0.50118}%
\definecolor{mygreen}{rgb}{0.00000,0.5,0.0}%
\definecolor{emerald}{rgb}{0.31, 0.78, 0.47}
\definecolor{burntorange}{rgb}{0.8, 0.33, 0.0}
\definecolor{darkmagenta}{rgb}{0.55, 0.0, 0.55}
\definecolor{deepmagenta}{rgb}{0.8, 0.0, 0.8}
\definecolor{forestgreen}{rgb}{0.0, 0.27, 0.13}
\pgfplotsset{every tick label/.append style={font=\scriptsize}}
\begin{tikzpicture}
\pgfplotsset{
    scale only axis,
    xmin=15, xmax=21,
    width=0.65\wlength,
height=0.9\hlength,
at={(0\wlength,0\hlength)},
}

\begin{axis}[
ymin=-24, ymax=-15,
xlabel style={font=\footnotesize \color{white!15!black},at={(0.5,-2ex)}},
xlabel={SNR [dB]},
ylabel style={font=\footnotesize \color{white!15!black},at={(-3ex,0.7)}, text width=2.5cm},
ylabel={NMSE [dB]},
xmajorgrids,
ymajorgrids,
ytick = {-22, -20, -18, -16},
yticklabels = {-22, -20, -18, -16},
xtick = {15, 16, 17, 18, 19, 20, 21},
xticklabels = {15, 16, 17, 18, 19, 20, 21},
]
\addplot [color=black, line width=1.0pt, mark size=1.0pt, mark=square*, mark options={solid, fill=myblack, draw=black}]
  table[row sep=crcr]{%
15	 	 	-17.64	 	 	 \\
15.33	 	 	-17.93	 	 	 \\
15.66	 	 	-18.2	 	 	 \\
16	 	 	-18.54	 	 	 \\
16.33	 	 	-18.8	 	 	 \\
16.66	 	 	-19.08	 	 	 \\
17	 	 	-19.4	 	 	 \\
17.33	 	 	-19.68	 	 	 \\
17.66	 	 	-19.97	 	 	 \\
18	 	 	-20.27	 	 	 \\
18.33	 	 	-20.49	 	 	 \\
18.66	 	 	-20.81	 	 	 \\
19	 	 	-21.05	 	 	 \\
19.33	 	 	-21.34	 	 	 \\
19.66	 	 	-21.57	 	 	 \\
20	 	 	-21.86	 	 	 \\
20.33	 	 	-22.11	 	 	 \\
20.66	 	 	-22.33	 	 	 \\
21	 	 	-22.56	 	 	 \\
21.33	 	 	-22.83	 	 	 \\
21.66	 	 	-23.08	 	 	 \\
22	 	 	-23.26	 	 	 \\
22.33	 	 	-23.53	 	 	 \\
22.66	 	 	-23.73	 	 	 \\
23	 	 	-23.95	 	 	 \\
24	 	 	-24.54	 	 	 \\
25	 	 	-25.1	 	 	 \\
30	 	 	-26.93	 	 	 \\
35	 	 	-27.78	 	 	 \\
};
\addplot [color=burntorange, line width=1.0pt, mark size=1.0pt, mark=diamond*, mark options={solid, fill=orange, draw=burntorange}]
  table[row sep=crcr]{%
15	-17.64	 \\
15.33	-18	 \\
15.66	-18.35	 \\
16	-18.65	 \\
16.33	-18.99	 \\
16.66	-19.32	 \\
17	-19.69	 \\
17.33	-20.05	 \\
17.66	-20.35	 \\
18	-20.67	 \\
18.33	-21.02	 \\
18.66	-21.38	 \\
19	-21.68	 \\
19.33	-22.03	 \\
19.66	-22.38	 \\
20	-22.66	 \\
20.33	-23.01	 \\
20.66	-23.36	 \\
21	-23.66	 \\
21.33	-24.01	 \\
21.66	-24.3	 \\
22	-24.64	 \\
22.33	-24.96	 \\
22.66	-25.25	 \\
23	-25.58	 \\
24	-26.52	 \\
25	-27.42	 \\
30	-31.32	 \\
35	-33.97	 \\
};
\addplot [color=myblue, line width=1.0pt, mark size=1.0pt, mark=*, mark options={solid, fill=mybluef, draw= myblue}]
table[row sep=crcr]{%
15 	 	 	 -17.96 	 	 	 \\
15.33 	 	 	 -18.26 	 	 	 \\
15.66 	 	 	 -18.57 	 	 	 \\
16 	 	 	 -18.93 	 	 	 \\
16.33 	 	 	 -19.26 	 	 	 \\
16.66 	 	 	 -19.59 	 	 	 \\
17 	 	 	 -19.86 	 	 	 \\
17.33 	 	 	 -20.21 	 	 	 \\
17.66 	 	 	 -20.54 	 	 	 \\
18 	 	 	 -20.87 	 	 	 \\
18.33 	 	 	 -21.18 	 	 	 \\
18.66 	 	 	 -21.51 	 	 	 \\
19 	 	 	 -21.82 	 	 	 \\
19.33 	 	 	 -22.11 	 	 	 \\
19.66 	 	 	 -22.43 	 	 	 \\
20 	 	 	 -22.79 	 	 	 \\
20.33 	 	 	 -23.09 	 	 	 \\
20.66 	 	 	 -23.4 	 	 	 \\
21 	 	 	 -23.71 	 	 	 \\
21.33 	 	 	 -24.04 	 	 	 \\
21.66 	 	 	 -24.35 	 	 	 \\
22 	 	 	 -24.68 	 	 	 \\
22.33 	 	 	 -24.95 	 	 	 \\
22.66 	 	 	 -25.26 	 	 	 \\
23 	 	 	 -25.61 	 	 	 \\
24 	 	 	 -26.45 	 	 	 \\
25 	 	 	 -27.31 	 	 	 \\
30 	 	 	 -31.45 	 	 	 \\
35 	 	 	 -34.24 	 	 	 \\
};
\end{axis}
\node[] at (2,2.65) {\footnotesize{(a) Weakly nonlinear (BO = 7 dB)}};
\end{tikzpicture}%
  \end{minipage}
  \begin{minipage}[b]{0.33\textwidth}
\definecolor{mycolor1}{rgb}{0.31373,0.31373,0.31373}%
\definecolor{mycolor2}{rgb}{0.00000,0.44706,0.74118}%
\definecolor{mycolor3}{rgb}{0.72941,0.83137,0.95686}%
\definecolor{myblack}{rgb}{0.31373,0.31373,0.31373}%
\definecolor{mybluef}{rgb}{0.00000,0.44706,0.74118}%
\definecolor{myblue}{rgb}{0.00000,0.30706,0.50118}%
\definecolor{mygreen}{rgb}{0.00000,0.5,0.0}%
\definecolor{emerald}{rgb}{0.31, 0.78, 0.47}
\definecolor{burntorange}{rgb}{0.8, 0.33, 0.0}
\definecolor{darkmagenta}{rgb}{0.55, 0.0, 0.55}
\definecolor{deepmagenta}{rgb}{0.8, 0.0, 0.8}
\definecolor{forestgreen}{rgb}{0.0, 0.27, 0.13}
\pgfplotsset{every tick label/.append style={font=\scriptsize}}
\begin{tikzpicture}
\pgfplotsset{
    scale only axis,
    xmin=15, xmax=21,
    width=0.65\wlength,
height=0.9\hlength,
at={(0\wlength,0\hlength)},
}

\begin{axis}[
ymin=-24, ymax=-15,
xlabel style={font=\footnotesize \color{white!15!black},at={(0.5,-2ex)}},
xlabel={SNR [dB]},
ylabel style={font=\footnotesize \color{white!15!black},at={(-3ex,0.7)}, text width=2.5cm},
ylabel={NMSE [dB]},
xmajorgrids,
ymajorgrids,
ytick = {-24, -22, -20, -18, -16},
yticklabels = {-24, -22, -20, -18, -16},
xtick = {15, 16, 17, 18, 19, 20, 21},
xticklabels = {15, 16, 17, 18, 19, 20, 21},
legend columns=4,
legend style={at={(-0.0,1.01)},
anchor=south west, legend cell align=left, align=left, draw=white!15!black}
]
\addplot [color=black, line width=1.0pt, mark size=1.0pt, mark=square*, mark options={solid, fill=myblack, draw=black}]
  table[row sep=crcr]{%
15	 	 	-17.06	 	 	 \\
15.33	 	 	-17.33	 	 	 \\
15.66	 	 	-17.6	 	 	 \\
16	 	 	-17.87	 	 	 \\
16.33	 	 	-18.06	 	 	 \\
16.66	 	 	-18.39	 	 	 \\
17	 	 	-18.57	 	 	 \\
17.33	 	 	-18.81	 	 	 \\
17.66	 	 	-19.09	 	 	 \\
18	 	 	-19.27	 	 	 \\
18.33	 	 	-19.5	 	 	 \\
18.66	 	 	-19.73	 	 	 \\
19	 	 	-19.92	 	 	 \\
19.33	 	 	-20.15	 	 	 \\
19.66	 	 	-20.35	 	 	 \\
20	 	 	-20.52	 	 	 \\
20.33	 	 	-20.76	 	 	 \\
20.66	 	 	-20.87	 	 	 \\
21	 	 	-21.06	 	 	 \\
21.33	 	 	-21.26	 	 	 \\
21.66	 	 	-21.39	 	 	 \\
22	 	 	-21.58	 	 	 \\
22.33	 	 	-21.71	 	 	 \\
22.66	 	 	-21.82	 	 	 \\
23	 	 	-21.97	 	 	 \\
24	 	 	-22.35	 	 	 \\
25	 	 	-22.63	 	 	 \\
30	 	 	-23.66	 	 	 \\
35	 	 	-24.05	 	 	 \\
};
\addlegendentry{\footnotesize Linear}
\addplot [color=burntorange, line width=1.0pt, mark size=1.0pt, mark=diamond*, mark options={solid, fill=orange, draw=burntorange}]
  table[row sep=crcr]{%
15	 	 	-17.43	 	 	 \\
15.33	 	 	-17.73	 	 	 \\
15.66	 	 	-18.08	 	 	 \\
16	 	 	-18.38	 	 	 \\
16.33	 	 	-18.73	 	 	 \\
16.66	 	 	-19.08	 	 	 \\
17	 	 	-19.35	 	 	 \\
17.33	 	 	-19.65	 	 	 \\
17.66	 	 	-20.01	 	 	 \\
18	 	 	-20.37	 	 	 \\
18.33	 	 	-20.65	 	 	 \\
18.66	 	 	-20.98	 	 	 \\
19	 	 	-21.31	 	 	 \\
19.33	 	 	-21.63	 	 	 \\
19.66	 	 	-21.95	 	 	 \\
20	 	 	-22.26	 	 	 \\
20.33	 	 	-22.55	 	 	 \\
20.66	 	 	-22.85	 	 	 \\
21	 	 	-23.14	 	 	 \\
21.33	 	 	-23.43	 	 	 \\
21.66	 	 	-23.7	 	 	 \\
22	 	 	-23.98	 	 	 \\
22.33	 	 	-24.27	 	 	 \\
22.66	 	 	-24.57	 	 	 \\
23	 	 	-24.83	 	 	 \\
24	 	 	-25.62	 	 	 \\
25	 	 	-26.27	 	 	 \\
30	 	 	-29.02	 	 	 \\
35	 	 	-30.38	 	 	 \\
};
\addlegendentry{\footnotesize Volterra ILA}
\addplot [color=myblue, line width=1.0pt, mark size=1.0pt, mark=*, mark options={solid, fill=mybluef, draw= myblue}]
table[row sep=crcr]{%
15 	 	 	 -17.9 	 	 	 \\
15.33 	 	 	 -18.2 	 	 	 \\
15.66 	 	 	 -18.54 	 	 	 \\
16 	 	 	 -18.83 	 	 	 \\
16.33 	 	 	 -19.15 	 	 	 \\
16.66 	 	 	 -19.47 	 	 	 \\
17 	 	 	 -19.81 	 	 	 \\
17.33 	 	 	 -20.13 	 	 	 \\
17.66 	 	 	 -20.4 	 	 	 \\
18 	 	 	 -20.74 	 	 	 \\
18.33 	 	 	 -21.05 	 	 	 \\
18.66 	 	 	 -21.35 	 	 	 \\
19 	 	 	 -21.67 	 	 	 \\
19.33 	 	 	 -21.95 	 	 	 \\
19.66 	 	 	 -22.27 	 	 	 \\
20 	 	 	 -22.58 	 	 	 \\
20.33 	 	 	 -22.88 	 	 	 \\
20.66 	 	 	 -23.21 	 	 	 \\
21 	 	 	 -23.5 	 	 	 \\
21.33 	 	 	 -23.75 	 	 	 \\
21.66 	 	 	 -24.08 	 	 	 \\
22 	 	 	 -24.38 	 	 	 \\
22.33 	 	 	 -24.66 	 	 	 \\
22.66 	 	 	 -24.9 	 	 	 \\
23 	 	 	 -25.21 	 	 	 \\
24 	 	 	 -26.1 	 	 	 \\
25 	 	 	 -26.92 	 	 	 \\
30 	 	 	 -30.27 	 	 	 \\
35 	 	 	 -32.45 	 	 	 \\
};
\addlegendentry{\footnotesize Volterra DLA}
\end{axis}
\node[] at (2, 2.6) {\footnotesize{(b) Mildly nonlinear (BO = 5 dB)}};
\end{tikzpicture}%
  \end{minipage}
  \begin{minipage}[b]{0.33\textwidth}
\definecolor{mycolor1}{rgb}{0.31373,0.31373,0.31373}%
\definecolor{mycolor2}{rgb}{0.00000,0.44706,0.74118}%
\definecolor{mycolor3}{rgb}{0.72941,0.83137,0.95686}%
\definecolor{myblack}{rgb}{0.31373,0.31373,0.31373}%
\definecolor{mybluef}{rgb}{0.00000,0.44706,0.74118}%
\definecolor{myblue}{rgb}{0.00000,0.30706,0.50118}%
\definecolor{mygreen}{rgb}{0.00000,0.5,0.0}%
\definecolor{emerald}{rgb}{0.31, 0.78, 0.47}
\definecolor{burntorange}{rgb}{0.8, 0.33, 0.0}
\definecolor{darkmagenta}{rgb}{0.55, 0.0, 0.55}
\definecolor{deepmagenta}{rgb}{0.8, 0.0, 0.8}
\definecolor{forestgreen}{rgb}{0.0, 0.27, 0.13}
\pgfplotsset{every tick label/.append style={font=\scriptsize}}
\begin{tikzpicture}
\pgfplotsset{
    scale only axis,
    xmin=15, xmax=21,
    width=0.65\wlength,
height=0.9\hlength,
at={(0\wlength,0\hlength)},
}

\begin{axis}[
ymin=-24, ymax=-15,
xlabel style={font=\footnotesize \color{white!15!black},at={(0.5,-2ex)}},
xlabel={SNR [dB]},
ylabel style={font=\footnotesize \color{white!15!black},at={(-3ex,0.7)}, text width=2.5cm},
ylabel={NMSE [dB]},
xmajorgrids,
ymajorgrids,
ytick = {-22, -20, -18, -16},
yticklabels = {-22, -20, -18, -16},
xtick = {15, 16, 17, 18, 19, 20, 21},
xticklabels = {15, 16, 17, 18, 19, 20, 21},
]
\addplot [color=black, line width=1.0pt, mark size=1.0pt, mark=square*, mark options={solid, fill=myblack, draw=black}]
  table[row sep=crcr]{%
15	 	 	-16.03	 	 	 \\
15.33	 	 	-16.19	 	 	 \\
15.66	 	 	-16.42	 	 	 \\
16	 	 	-16.61	 	 	 \\
16.33	 	 	-16.81	 	 	 \\
16.66	 	 	-17.01	 	 	 \\
17	 	 	-17.19	 	 	 \\
17.33	 	 	-17.39	 	 	 \\
17.66	 	 	-17.52	 	 	 \\
18	 	 	-17.7	 	 	 \\
18.33	 	 	-17.83	 	 	 \\
18.66	 	 	-18.01	 	 	 \\
19	 	 	-18.08	 	 	 \\
19.33	 	 	-18.26	 	 	 \\
19.66	 	 	-18.4	 	 	 \\
20	 	 	-18.49	 	 	 \\
20.33	 	 	-18.62	 	 	 \\
20.66	 	 	-18.77	 	 	 \\
21	 	 	-18.83	 	 	 \\
21.33	 	 	-18.96	 	 	 \\
21.66	 	 	-19	 	 	 \\
22	 	 	-19.17	 	 	 \\
22.33	 	 	-19.21	 	 	 \\
22.66	 	 	-19.31	 	 	 \\
23	 	 	-19.35	 	 	 \\
24	 	 	-19.58	 	 	 \\
25	 	 	-19.72	 	 	 \\
30	 	 	-20.25	 	 	 \\
35	 	 	-20.4	 	 	 \\
};
\addplot [color=burntorange, line width=1.0pt, mark size=1.0pt, mark=diamond*, mark options={solid, fill=orange, draw=burntorange}]
  table[row sep=crcr]{%
15	 	 	-16.84	 	 	 \\
15.33	 	 	-17.2	 	 	 \\
15.66	 	 	-17.44	 	 	 \\
16	 	 	-17.72	 	 	 \\
16.33	 	 	-18.03	 	 	 \\
16.66	 	 	-18.32	 	 	 \\
17	 	 	-18.63	 	 	 \\
17.33	 	 	-18.92	 	 	 \\
17.66	 	 	-19.15	 	 	 \\
18	 	 	-19.47	 	 	 \\
18.33	 	 	-19.74	 	 	 \\
18.66	 	 	-19.94	 	 	 \\
19	 	 	-20.22	 	 	 \\
19.33	 	 	-20.5	 	 	 \\
19.66	 	 	-20.72	 	 	 \\
20	 	 	-20.98	 	 	 \\
20.33	 	 	-21.22	 	 	 \\
20.66	 	 	-21.49	 	 	 \\
21	 	 	-21.64	 	 	 \\
21.33	 	 	-21.91	 	 	 \\
21.66	 	 	-22.06	 	 	 \\
22	 	 	-22.33	 	 	 \\
22.33	 	 	-22.5	 	 	 \\
22.66	 	 	-22.68	 	 	 \\
23	 	 	-22.84	 	 	 \\
24	 	 	-23.15	 	 	 \\
25	 	 	-23.6	 	 	 \\
30	 	 	-24.68	 	 	 \\
35	 	 	-25.04	 	 	 \\
};
\addplot [color=myblue, line width=1.0pt, mark size=1.0pt, mark=*, mark options={solid, fill=mybluef, draw= myblue}]
table[row sep=crcr]{%
15	 	 	-17.72	 	 	 \\
15.33	 	 	-18	 	 	 \\
15.66	 	 	-18.38	 	 	 \\
16	 	 	-18.65	 	 	 \\
16.33	 	 	-18.93	 	 	 \\
16.66	 	 	-19.24	 	 	 \\
17	 	 	-19.56	 	 	 \\
17.33	 	 	-19.86	 	 	 \\
17.66	 	 	-20.15	 	 	 \\
18	 	 	-20.45	 	 	 \\
18.33	 	 	-20.74	 	 	 \\
18.66	 	 	-21.01	 	 	 \\
19	 	 	-21.34	 	 	 \\
19.33	 	 	-21.62	 	 	 \\
19.66	 	 	-21.91	 	 	 \\
20	 	 	-22.2	 	 	 \\
20.33	 	 	-22.47	 	 	 \\
20.66	 	 	-22.7	 	 	 \\
21	 	 	-23.03	 	 	 \\
21.33	 	 	-23.3	 	 	 \\
21.66	 	 	-23.52	 	 	 \\
22	 	 	-23.8	 	 	 \\
22.33	 	 	-24.1	 	 	 \\
22.66	 	 	-24.3	 	 	 \\
23	 	 	-24.58	 	 	 \\
24	 	 	-25.23	 	 	 \\
25	 	 	-25.97	 	 	 \\
30	 	 	-28.52	 	 	 \\
35	 	 	-29.81	 	 	 \\
};
\end{axis}
\node[] at (2, 2.6) {\footnotesize{(c) Strongly nonlinear (BO = 3 dB)}};
\end{tikzpicture}%
  \end{minipage}  
\vspace{-1.2cm}
\end{figure}
We further plot the GMI performance of each DPDs for the three different cases in Fig.~\ref{fig:snr_vs_gmi_16qam}(d). Unlike other DPDs, we see that GMI degradation for the Volterra-DLA DPD is negligible when nonlinearity is boosted up. At 18 dB SNR, the Volterra-ILA achieves GMI of 5.71 bits and 5.82 bits for back-off of 5 dB and 3 dB, respectively. These value for the Volterra-DLA are 5.87 bits and 5.85 bits respectively.
In Fig.~\ref{fig:snr_vs_gmi_16qam}(e) and (f), we show the histograms of the real part of the received symbols for Volterra-ILA and Volterra-DLA, respectively. We see that with Volterra-DLA Fig.~\ref{fig:snr_vs_gmi_16qam}(f), the signal levels are intact and not spread in contrast to the case with Volterra-ILA Fig.~\ref{fig:snr_vs_gmi_16qam}(e).

\section{Conclusion}
In this paper, we presented a simplified method for direct learning of Volterra series-based DPD. The method could also be applied to other DPDs such as memory polynomials. Using simulations, we showed the superior performance of our learning method in comparison to that of ILA providing around 1 dB extra NMSE gain.

\textit{This work was carried out under the EU Marie Sk{\l}odowska-Curie project FONTE (No. 766115)}.

{
\small
\bibliography{sample}
}

\end{document}